\documentclass[prl,aps,showpacs,twocolumn,floatfix]{revtex4}
\usepackage{amsmath,amsfonts,amssymb,graphicx}
\newcommand{\cd}{c^\dag}
\newcommand{\bd}{b^\dag}
\newcommand{\vac}{{\vert0\rangle}}
\newcommand{\hh}{
        \begin{picture}(9,8)(-1,1)
         \put(0,0){\line(1,0){7}}
         \put(7,7){\line(-1,0){7}}
         \put(0,0){\circle*{3}}
         \put(0,7){\circle*{3}}
         \put(7,0){\circle*{3}}
         \put(7,7){\circle*{3}}
        \end{picture}
}
\newcommand{\vv}{
        \begin{picture}(9,8)(-1,1)
         \put(0,0){\line(0,1){7}}
         \put(7,7){\line(0,-1){7}}
         \put(0,0){\circle*{3}}
         \put(0,7){\circle*{3}}
         \put(7,0){\circle*{3}}
         \put(7,7){\circle*{3}}
        \end{picture}
}
\begin{document}
\title{Superconductivity in strongly repulsive fermions: the role of
       kinetic-energy frustration}
\author{L. Isaev$^1$}
\author{G. Ortiz$^1$}
\author{C. D. Batista$^2$}
\affiliation{$^1$Department of Physics, Indiana University,
                 Bloomington IN 47405 \\
             $^2$Theoretical Division, Los Alamos National Laboratory,
                 Los Alamos NM 87545}
\begin{abstract}
 We discuss a physical mechanism of a non-BCS nature which can stabilize a
 superconducting state in a {\it strongly repulsive} electronic system. By
 considering the two-dimensional Hubbard model with spatially modulated
 electron hoppings, we demonstrate how kinetic-energy frustration can lead to
 robust $d$-wave superconductivity at {\it arbitrarily} large on-site
 repulsion. This phenomenon should be observable in experiments using fermionic
 atoms, e.g. ${}^{40}K$, in specially prepared optical lattices.
\end{abstract}
\pacs{74.20.Mn, 71.10.Fd}
\maketitle

{\it Introduction.--}
One of the long-standing fundamental questions in condensed-matter physics is
whether it is possible to realize a superconducting (SC) state in a system
consisting only of electrons subject to a strong Coulomb repulsion, and if so,
what is the minimal set of necessary physical assumptions. An early attempt to
provide an answer was made by Kohn and Luttinger \cite{Kohn_1965}, who proposed
a weak-coupling BCS-like mechanism. While their idea was never confirmed
experimentally, there exist numerous {\it strongly} correlated systems whose SC
behavior occurs without any obvious pairing glue, such as phonons, between the
electrons. Examples are high-$T_c$ cuprates and heavy fermion compounds. The
current consensus is that superconductivity in these materials has an
unconventional, i.e. non-BCS, character \cite{Monthoux_2007}. Understanding the
microscopic origin of this intriguing phenomenon remains a challenge. Here we
address the above question by performing a controlled derivation of the SC
ground state (GS) for a {\it strongly-repulsive} Hubbard model with spatially
modulated transfer integrals.

One possible way of stabilizing a Cooper pair condensate in a repulsive system
is to introduce microscopic inhomogeneities. Indeed, the nanoscale spin and
charge modulations, observed in scattering \cite{Tranquada_1995}, ARPES
\cite{Valla_2006} and STM \cite{Wise_2009} experiments, seem to be ubiquitous
in high-$T_c$ materials \cite{Dagotto_2005} and often accompany the emergence
of the SC state. Theoretically it has been argued that these inhomogeneities
are quite relevant for the superconductivity \cite{Eroles_2000,Kivelson_2007}
and, in fact, seem to assist the Cooper pairing. This was demonstrated in
\cite{Eroles_2000,Tsai_2008} by using exact diagonalization of strongly
interacting models in finite lattices. In Ref. \onlinecite{Yao_2007} the
authors studied the Hubbard model on a checkerboard lattice, composed of weakly
coupled $2\times2$ plaquettes, and showed that the SC phase can be stabilized
in a relatively narrow interval of the on-site repulsion $U$. Earlier, a
similar problem was considered in \cite{Barabanov_1989}. Another ingredient,
whose importance for superconductivity was largely overlooked, is the range of
the transfer integrals beyond nearest-neighbors (NN). The next-NN (NNN)
hopping, $t^\prime$, was shown to enhance $d_{x^2-y^2}$-like pairing
correlations in the $t$-$t^\prime$-$J$ model on finite clusters
\cite{Martins_2001}. Physically, its main qualitative effect is the possible
{\it frustration} of the kinetic-energy term: the smallest closed paths in the
lattice are triangles instead of squares.

\begin{figure}[t]
 \begin{center}
  \includegraphics[width=0.64\columnwidth]{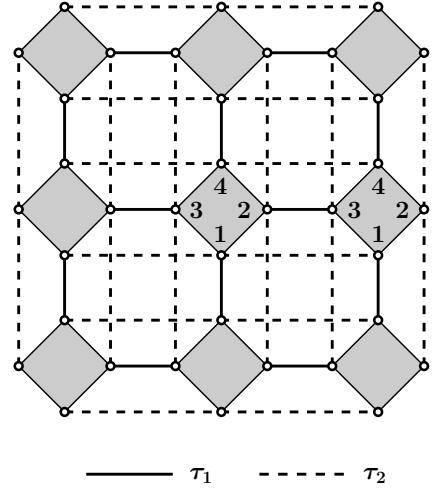}
 \end{center}
 \caption{The tetrahedral lattice topology. The interplaquette hopping
          amplitudes are: NN $\tau_1$ (solid lines) and NNN $\tau_2$ (dashed
          lines). We only consider the case $\tau_2\leqslant\tau_1$.}
 \label{fig_lattice}
\end{figure}

In the present Letter we explicitly demonstrate how {\it local} kinetic-energy
{\it frustration} can stabilize the SC state in a {\it strongly repulsive}
two-dimensional Hubbard model. The lattice, on which the model is defined, is
presented in Fig. \ref{fig_lattice}. It consists of weakly-coupled tetrahedra,
i.e. plaquettes with frustrated hoppings along the diagonals. We show that a
$d_{x^2-y^2}$--wave SC phase exists for {\it arbitrarily} strong repulsion $U$.
In fact, the problem can be treated analytically in the strong-coupling regime.

Our motivation to study this system is not purely academic. Advances in
experimental methods of preparation and manipulation of ultracold fermion atoms
in optical lattices provide a controlled way of testing the above-mentioned
theoretical ideas. For example, in recent experiments
\cite{Jordens_2008,Schneider_2008} the observation of a Mott state with
${}^{40}K$ atoms was reported. Moreover, an experiment aimed to find $d$-wave
superconductivity in a checkerboard Hubbard model was proposed in Ref.
\onlinecite{Rey_2009}.

{\it Model.--}
Let us consider the repulsive Hubbard model:
\begin{equation}
 H=-\sum_{\langle ij\rangle,\sigma}t_{ij}\bigl(\cd_{i\sigma}c_{j\sigma}+
 {\rm h.c.}\bigr)+U\sum_in^e_{i\uparrow}n^e_{i\downarrow},
 \label{hub_mod}
\end{equation}
defined on the lattice, Fig. \ref{fig_lattice}, in terms of fermionic
(creation) operators $\cd_{i\sigma}$. Here $\langle ij\rangle$ denotes links
connecting sites $i$ and $j$, $\sigma=\{\uparrow,\downarrow\}$ is the electron
spin, and $n^e_{i\sigma}=\cd_{i\sigma}c_{i\sigma}$. The amplitudes $t_{ij}$
take four possible values: (i) $t$ for links $\langle12\rangle$,
$\langle13\rangle$, $\langle24\rangle$ and $\langle34\rangle$; (ii) $t^\prime$
for the diagonals $\langle14\rangle$ and $\langle23\rangle$; (iii) $\tau_1$ for
NN links, connecting two plaquettes; (iv) $\tau_2$ for NNN interplaquette
links.

We will consider the case $\tau_{1,2}\ll t$, $t^\prime$, $U$, which allows for
a controlled perturbative expansion of the Hamiltonian \eqref{hub_mod}. To
demonstrate the existence of a robust SC phase, we derive a low-energy
effective model, accurate to second order in $\tau_{1,2}$. In general, this is
doable only numerically. However, in the limit $t$, $t^\prime\ll U$, we can
keep only lowest-order terms in $t_{ij}/U$, and thus provide a closed form for
the effective Hamiltonian (EH). The stability of the Cooper pair condensate can
be tuned by changing the ratio $t^\prime/t$. There is an ``optimal'' value of
this ratio, which ensures a finite energy gap (hole binding energy) between the
plaquette states with one and two holes, for all finite $U$.

\begin{figure}[t]
 \begin{center}
  \includegraphics[width=\columnwidth]{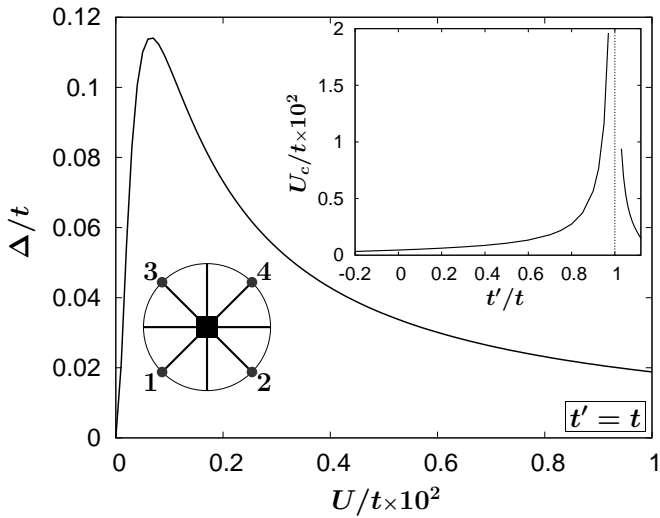}
 \end{center}
 \caption{Hole binding energy $\Delta(U)$. The asymptotic behavior is:
          $\Delta(U\gg t)\approx2t^2/U$ and $\Delta(U\ll t)\approx U^2/32t$.
          Upper inset: critical value $U_c(t^\prime)$ [$\Delta(U_c)=0$]. Lower
          inset: group $C_{4v}$. Numbers indicate plaquette vertices. The black
          square denotes four-fold axis $C_4$, horizontal and vertical lines --
          primary symmetry planes $\sigma_v$, diagonals -- secondary planes
          $\sigma^\prime_v$.}
 \label{fig_gap}
\end{figure}

{\it Single-plaquette states.--}
The Hubbard Hamiltonian on a single plaquette can be diagonalized exactly
\cite{Barabanov_1989} by using representations of the crystallographic group
$C_{4v}$ (see the lower inset in Fig. \ref{fig_gap}). As a result, we can
determine the hole binding energy
$\Delta=2\epsilon_0(3)-\epsilon_0(2)-\epsilon_0(4)$, where $\epsilon_0(N_e)$ is
the GS for a given number of electrons $N_e$. Positive values of $\Delta$
correspond to binding of two holes. In general, $\Delta$ is positive only in a
finite range of $U$. At some critical value $U_c(t^\prime)$, shown in the upper
inset of Fig. \ref{fig_gap}, it changes sign and remains negative as
$U\to\infty$. There is a special ratio, $t^\prime/t=1$, at which $U_c$ diverges
and $\Delta$ stays {\it positive} for any value of $U$ (see the main panel of
Fig. \ref{fig_gap}). This results from the maximal frustration of the
single-hole kinetic energy. The GS energy for 4 electrons (zero holes),
$\epsilon_0(4)\to0$ for $U\to\infty$ because the particles cannot move. On the
other hand, in this limit $\epsilon_0(2)=2\epsilon_0(3)$, which means that
there is no kinetic-energy gain for creating two holes on different plaquettes;
i.e., the single-hole kinetic energy is optimally frustrated. The exchange
interaction $J=4t^2/U$, that appears for finite $t/U\ll 1$, leads to pairing
($\Delta>0$) because the magnetic configuration of two plaquettes with one hole
in each of them is more frustrated than the configuration with two holes in the
same plaquette. This leads to a positive value of $\Delta=J/2$.

From now on we will only consider the maximally frustrated point
$t^\prime/t=1$. Then, the symmetry group ${\cal G}$ of the single-plaquette
Hamiltonian is larger than $C_{4v}$ (symmetry group for arbitrary
$t^\prime/t$), and contains all the independent permutations of any pair of
vertices of the plaquette. This symmetry translates into a GS degeneracy at
half-filling. There are two $SU(2)$-singlet states: one transforming as the
identity representation of $C_{4v}$, $A_1$ ($s$-wave), and the other -- as
$B_1$ ($d_{x^2-y^2}$-wave) \cite{Landau_QM}. These states are connected by
symmetry operations from the factor group ${\cal G}/C_{4v}$. The two-electron
GS is also a singlet and belongs to the identity representation of ${\cal G}$.
The $N_e=3$ GS has $S=1/2$ and is six-fold degenerate.

General expressions for these eigenstates are quite cumbersome. However, to the
lowest order in $t/U$, we can consider only states without doubly occupied
sites. Hence, we have the GS for $N_e=2$:
$\vert\Omega_2\rangle=\bigl(1/2\sqrt{3}\bigr)\sum_{\langle ij\rangle}
s^\dag_{ij}\vac$
with the summation extended over all links of a plaquette; and for $N_e=4$:
$\vert\Omega^{s,d}_4\rangle={\cal N}_{s,d}\bigl(s^\dag_{13}s^\dag_{24}\pm
s^\dag_{12}s^\dag_{34}\bigr)\vac$. In these expressions $s^\dag_{ij}$ is a
singlet creation operator,
$s^\dag_{ij}=\cd_{i\uparrow}\cd_{j\downarrow}-
\cd_{i\downarrow}\cd_{j\uparrow}$, $\vac$ is the empty state and
${\cal N}_s=-1/2$, ${\cal N}_d=1/2\sqrt{3}$. Finally, we introduce operators
$P_{ij}$, which permute sites $i$ and $j$. In the basis
$\{\vert\Omega^{s}_4\rangle,\vert\Omega^{d}_4\rangle \}$, $P_{12}$ and $P_{13}$
have the form:
$P_{12,13}=-\sigma^z/2\pm\sqrt{3}\sigma^x/2$ with $\sigma^\alpha$
($\alpha=x,z$) Pauli matrices. We will use this expression to determine
symmetries of the effective model.

{\it Effective low-energy model.--}
The low-energy spectrum of decoupled plaquettes has a gap $\Delta$ to
single-hole ($N_e=3$ on each plaquette) states. Here we consider the effect of
finite hopping amplitudes $\tau_{1,2}$ by assuming that
$0\leqslant\tau_{1,2}\ll\Delta\sim t^2/U\ll t\ll U$. The second inequality
allows us to treat interplaquette hoppings perturbatively. The fourth one
allows us to exclude states with doubly occupied sites, i.e. use as a basis the
states $\vert\Omega_2\rangle$ and $\vert\Omega^{s,d}_4\rangle$. Finally, the
third inequality constrains the choice of the virtual states: only states that
belong to the $N_e=3$ GS sextet contribute to lowest order. We will also assume
that $\tau_2\leqslant\tau_1$.

The second-order EH can be symbolically written as:
\begin{displaymath}
 H_{\rm eff}={\cal P}^{(0)}H_\tau\bigl(1-{\cal P}^{(0)}\bigr)
 \frac{1}{E_0-H^{(0)}}\bigl(1-{\cal P}^{(0)}\bigr)H_\tau{\cal P}^{(0)},
\end{displaymath}
where $H^{(0)}$ describes a set of noninteracting plaquettes in
\eqref{hub_mod}, $E_0$ is its GS energy, $H_\tau$ denotes plaquette
interactions, and ${\cal P}^{(0)}$ is a projector onto the subspace with
$N_e=2$ or $4$ on each plaquette. Next, we associate the product of the
two-electron plaquette GS with the vacuum:
$\vert{\rm vac}\rangle=\prod_x\vert\Omega_2\rangle_x$ and each member of the
four-electron GS doublet on plaquette $x$ -- with a hard-core boson:
$\vert\Omega_4^\alpha\rangle_x=\bd_{x\alpha}\vert\Omega_2\rangle_x$, where
$\alpha=s$ or $d$ represents the pseudospin index. The algebra generated by
$b_{x\sigma}$ was discussed in Ref. \onlinecite{Batista_2002}. Thus, the
effective low-energy theory, given by $H_{\rm eff}$, describes a system of
two-flavor hard-core bosons, propagating in the coarse-grained plaquette
lattice of Fig. \ref{fig_lattice}. In terms of these boson operators we have:
\begin{align}
 H_{\rm eff}=&\sum_{\langle xy\rangle,\alpha\beta}t^{\rm eff}_{\alpha\beta}
 \bigl(\bd_{x\alpha}b_{y\beta}+\bd_{y\beta}b_{x\alpha}\bigr)-\mu\sum_{x,\alpha}
 n_{x\alpha}+ \label{eff_model} \\
 &+\sum_{\langle xy\rangle,\alpha\beta}V^{\rm eff}_{\alpha\beta}\bigl[(1-n_x)
 \bd_{y\alpha}b_{y\beta}+(1-n_y)\bd_{x\alpha}b_{x\beta}\bigr], \nonumber
\end{align}
where $\langle xy\rangle$ denotes NN plaquettes, $t^{\rm eff}_{\alpha\beta}$
are corresponding hopping amplitudes, $V^{\rm eff}_{\alpha\beta}$ --
density-density and local spin-flip interactions,
$n_{x\alpha}=\bd_{x\alpha}b_{x\alpha}$ and $n_x=n_{xs}+n_{xd}$, and $\mu$ is
the chemical potential. Direct interactions between pseudospins, like
Heisenberg terms, are not present to lowest order in $\Delta/t$ and $t/U$.

Some general properties of $t^{\rm eff}$ and $V^{\rm eff}$ can be established
by symmetry arguments. First, the Hamiltonian \eqref{hub_mod} is invariant
under reflections in the planes which include $\tau_1$ links, e.g. the plane
connecting sites $2$ and $3$ in Fig. \ref{fig_lattice}. The states
$\vert\Omega_2\rangle$ and $\vert\Omega_4^s\rangle$ are symmetric under this
operation, while the $d$-wave state $\vert\Omega_4^d\rangle$ is antisymmetric.
Consequently, the off-diagonal matrix elements of $t^{\rm eff}_{sd}$ and
$V^{\rm eff}_{sd}$ vanish:
$t^{\rm eff}_{\alpha\beta}=t^{\rm eff}_{\alpha\alpha}\delta_{\alpha\beta}$,
$V^{\rm eff}_{\alpha\beta}=V^{\rm eff}_{\alpha\alpha}\delta_{\alpha\beta}$.
This result is independent of the assumptions made regarding the relative
magnitude of $\tau_{1,2}$, $U$ and $t$.

Another observation concerns the diagonal elements of $t^{\rm eff}$ and
$V^{\rm eff}$ in the special cases $\tau_2=\tau_1$ and $\tau_2=0$. In the
first case, we consider the two plaquettes with numbered sites, shown in Fig.
\ref{fig_lattice}, and perform simultaneous permutations of vertices
$1\leftrightarrow2$ on the left plaquette and $1\leftrightarrow3$ on the right
one. Each operation is a symmetry of the single-plaquette Hamiltonian. Their
combination amounts to interchanging the $\tau_1$ and $\tau_2$ links, which is
now a symmetry of the connecting Hamiltonian. Using the relation
$P_{12,13}\vert\Omega_2\rangle=\vert\Omega_2\rangle$, it is easy to show that
for $\tau_1=\tau_2$: $t^{\rm eff}_{ss}=-t^{\rm eff}_{dd}$ and
$V^{\rm eff}_{ss}=V^{\rm eff}_{dd}$. In the case $\tau_2=0$, when the
plaquettes are connected by only one $\tau_1$ link, the second-order virtual
hopping of an electron can only proceed through an intermediate state, whose
energy is of order $U$. Therefore, in the approximation formulated above,
$t^{\rm eff}$ must vanish. On the contrary, $V^{\rm eff}$ is not associated
with the net electron transfer and remains finite.

In general, a direct calculation yields the precise form of the coefficients
$t^{\rm eff}$ and $V^{\rm eff}$: 
\begin{align}
 t^{\rm eff}_{\alpha\beta}=&-\bigl(\tau_1^2/6\Delta\bigr){\rm diag}\bigl\{
 r_\tau(2r_\tau+1),\,\,\,-3r_\tau\bigr\}; \label{eff_coeff}  \\
 V^{\rm eff}_{\alpha\beta}=&-\bigl(\tau_1^2/48\Delta\bigr){\rm diag}\bigl\{
 9+8r_\tau+16r_\tau^2,\,\,\,9+24r_\tau^2\bigr\} \nonumber
\end{align}
with $r_\tau=\tau_2/\tau_1$. Clearly, in the two special cases, discussed above
-- $r_\tau=1$ and $0$ -- the EH \eqref{eff_model} becomes pseudospin symmetric.
The second case is irrelevant for the purposes of studying the SC state, while
the first one, $r_\tau=1$, is quite instructive. Indeed, in this case we can
use the Perron-Frobenius theorem to prove that there exists a
pseudospin-polarized GS \cite{Fledderjohann_2005}. The Hamiltonian can then be
written only in terms of spinless bosons, say $b_{xd}$, and maps onto the
spin-$1/2$ XXZ model in a magnetic field $\mu$, via the Matsubara-Matsuda
transformation \cite{Batista_2004}. The phase diagram of this model contains
N\'eel, canted XY-antiferromagnetic and fully polarized states that are
immediately identified with the density-wave (DW), Bose-Einstein condensate
(BEC) of Cooper pairs, and Mott phases, respectively. The DW and BEC states are
separated by a 1st order quantum phase transition.

\begin{figure}[b]
 \begin{center}
  \includegraphics[width=\columnwidth]{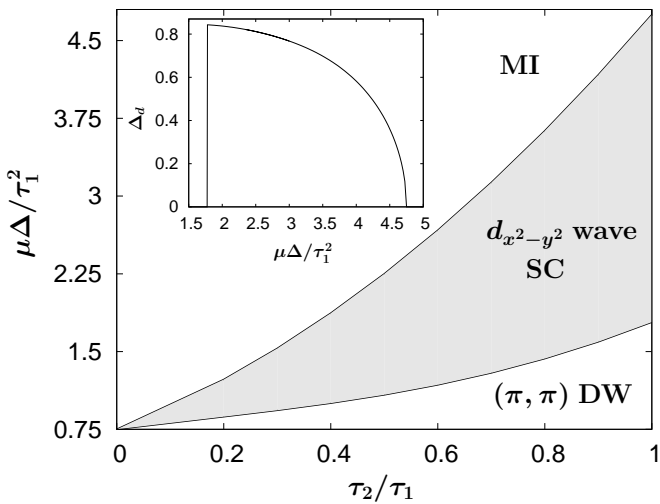}
 \end{center}
 \caption{Low-energy phase diagram of the Hubbard model Eq. \eqref{hub_mod}.
          Phases are: $s$-wave density wave (DW) with wavevector $(\pi,\pi)$;
          $d$-wave SC, which corresponds to a BEC of $b_d$; Mott insulating
          (MI) phase with $\langle n_x\rangle=1$. The DW -- SC phase transition
          is 1st order; the SC -- MI  transition is 2nd order. The inset shows
          $\Delta_d\sim\langle b_{xd}\rangle$ for $r_\tau=1$.}
 \label{fig_phase_diag}
\end{figure}

We do not expect the physics to change qualitatively for $0<r_\tau<1$. It is
known that the usual mean-field approximation yields satisfactory results for
$r_\tau=1$ when compared to Monte-Carlo simulations \cite{Batrouni_2000}. Thus,
we anticipate that the rest of the phase diagram, along the $r_\tau$ axis, can
be described within a simple variational approach. We employ the method of
\cite{Isaev_2009}, which includes short-range quantum fluctuations and, as a
limit, contains the semiclassical spin-wave ansatz. The resulting phase
diagram, obtained using $2\times2$ site clusters (in the plaquette lattice), is
presented in Fig. \ref{fig_phase_diag}. For any finite $0<r_\tau<1$ the system
exhibits the same three phases, as in the case $r_\tau=1$. The SC and DW phases
are again separated by a 1st order transition. The transition between SC and
Mott phases is 2nd order. In the Mott state there is exactly one boson per
site; i.e., the electron filling is $1/2$. In this phase the pseudospin
polarization is undefined, as the Hamiltonian \eqref{eff_model} becomes
spin-independent. Interestingly, the DW phase is of an $s$-wave nature, due to
the fact that the expectation value of the kinetic energy vanishes, while
density-density interactions favor the $s$-wave pseudospin polarization.

The SC state has a $d_{x^2-y^2}$-wave symmetry. The structure of the ``Cooper
pair'' can be determined by observing that
$b_d={\cal D}/3-(1/4)(s_{14}+s_{23})\bigl(\vert\vv\rangle\langle\vv\vert-\vert
\hh\rangle\langle\hh\vert\bigr)$, where
$\vert\vv\rangle=s^\dag_{13}s^\dag_{24}\vac$,
$\vert\hh\rangle=s^\dag_{12}s^\dag_{34}\vac$ and
${\cal D}=s_{13}+s_{24}-s_{12}-s_{34}$ (see the lower inset of Fig.
\ref{fig_gap}). Hence, despite the apparent complexity of the SC phase, it can
still be characterized by a familiar $d$-wave order parameter
$\Delta_d=\langle{\cal D}\rangle$, shown in the inset of Fig.
\ref{fig_phase_diag} for $r_\tau=1$. As $r_\tau$ decreases, the height of the
SC dome gradually diminishes and disappears at $r_\tau=0$. Thus, for any
$0<r_\tau<1$ there is an interval of $\mu$ where the SC phase is stabilized.
This conclusion becomes rigorous in the dilute limit of particles or holes, by
virtue of the inequality $\vert t^{\rm eff}_{dd}\vert>\vert t^{\rm
eff}_{ss}\vert$, valid for $r_\tau<1$ [see Eq. \eqref{eff_coeff}].

{\it Discussion.--}
Our phase diagram, Fig. \ref{fig_phase_diag}, was obtained in the
strong-coupling limit $U\gg t$, where one can derive the effective model of
Eqs. \eqref{eff_model}, \eqref{eff_coeff}. The EH becomes
increasingly complicated for intermediate couplings $U\sim t$, because of the
large number of virtual transitions. In this regime, the existence of $d$-wave
superconductivity in the nonfrustrated Hubbard model was argued in
\cite{Yao_2007} based on a first-order EH, treated within a mean-field
approximation, and in the weak-coupling regime $U\ll t$ in \cite{Raghu_2010}.
Therefore, we expect the SC phase to persist for $U\sim t$ in our frustrated
case as well. However, regardless of the magnitude of $U$, the SC state is
quite sensitive to the presence of longer-range repulsions. For instance, an
interaction of the form $\sum V_{ij}n^e_in^e_j$ with $V_{ij}=V$ for all links
within the plaquette, will suppress the local hole binding if
$V\geqslant V_c=0.114t$. For $V<V_c$ the SC phase is stable only in a finite
interval of $U$ around $U\sim7t$.

Our theory highlights the importance of the kinetic-energy frustration for
stabilizing the SC state. Locally, pairing competes against the kinetic energy
and can be increased by frustrating the latter. This principle guides the
choice of the {\it elementary unit}, e.g., tetrahedron. The {\it connectivity}
of the lattice, built from these blocks is another essential ingredient. Here
we used the lattice of Fig. \ref{fig_lattice} to demonstrate the existence of
the SC state in a physically transparent way. However, we also considered the
usual checkerboard lattice \cite{Yao_2007}. In this case the relation between
coefficients in the EH is such that the phase-separated state can suppress
superconductivity in a certain region of the phase diagram. The importance of
the lattice topology is further illustrated by the case $\tau_2=0$. Without the
interplaquette hopping \eqref{eff_coeff}, the global phase coherence can be
established only in higher orders in $1/U$, leading to a quite fragile SC
state.  

Finally, we believe that the lattice of Fig. \ref{fig_lattice} can be realized
using ideas of Refs. \onlinecite{Rey_2009,Jiang_2009}. Indeed, our effective
strong-coupling model can be easily extended to the currently experimentally
realizable regime $t'/t\lesssim0.5$ under the condition $t\ll U<U_c$, which can
still be fulfilled for $t'/t=0.5$ because $U_c\approx11t$ (see inset of Fig.
\ref{fig_gap}). The resulting phase diagram is qualitatively the same as the
one shown in Fig. \ref{fig_phase_diag}. Thus, results of the present Letter can
be tested in future cold atom experiments.

{\it Acknowledgements.--}
LI acknowledges the hospitality of CNLS at LANL. CDB was supported by US DOE.

\end{document}